\documentclass[10pt]{article}

\usepackage{amsmath}
\usepackage{amssymb}
\usepackage{graphicx}
\usepackage{cite}
\usepackage{color}
\usepackage{multirow}

\usepackage{setspace}
\usepackage[labelfont=bf,labelsep=period,justification=raggedright]{caption}

\makeatletter
\renewcommand{\@biblabel}[1]{\quad#1.}
\makeatother

\date{}

\begin{document}

\begin{flushleft}
{\Large
\textbf{Segregation in Religion Networks}
}
\\
Jiantao Hu$^{1}$,
Qian-Ming Zhang$^{2}$,
Tao Zhou$^{1,2,\S}$
\\
{\bf1} CompleX Lab, University of Electronic Science and Technology of China, Chengdu 611731, People's Republic of China
\\
{\bf2} Big Data Research Center, University of Electronic Science and Technology of China, Chengdu 611731, People's Republic of China
\\
Corresponding to: $\S$ zhutou@ustc.edu
\end{flushleft}

{\bf
Religious beliefs could facilitate human cooperation \cite{Sosis2003Signaling,Atkinsonab2011Beliefs,Xygalatas2013Extreme,Baumard2013Explaining,Botero2014The,Purzycki2016Moralistic}, promote civic engagement \cite{Wilson1997Who,Graham2010Beyond,Lewisa2013Religion,Power2017Social}, improve life satisfaction \cite{Lim2010Religion,Okulicz-Kozaryn2010,Ritter2014Happy} and even boom economic development \cite{Iannaccone1998Introduction,Barro2003Religion,Norenzayan2016Cultural}. On the other side, some aspects of religion may lead to regional violence, intergroup conflict and moral prejudice against atheists \cite{Appleby1999The,Atran2012Religious,Neuberg2014Religion,Edgell2006Atheists,Gervais2011Do,Gervais2014Everything,Gervais2017Global}. Analogous to the separation of races \cite{Lewis2013The}, the religious segregation is a major ingredient resulting in increasing alienation, misunderstanding, cultural conflict and even violence among believers of different faiths \cite{Atran2012Religious,Neuberg2014Religion,Kiernan1974Where}. Thus far, quantitative understanding of religious segregation is rare. Here we analyze a directed social network extracted from \emph{weibo.com} (the largest directed social network in China, similar to \emph{twitter.com}), which is consisted of 6875 believers in Christianism, Buddhism, Islam and Taoism. This religion network is highly segregative, with only 1.6\% of links connecting individuals in different religions. Comparative analysis shows that the extent of segregation for different religions is much higher than that for different races and slightly higher than that for different political parties. The few cross-religion links play a critical role in maintaining network connectivity, being remarkably more important than links with highest betweennesses \cite{Girvan2002Community} or bridgenesses \cite{Cheng2010Bridgeness}. Further content analysis shows that 46.7\% of these cross-religion links are probably related to charitable issues. Our findings provide quantitative insights into religious segregation and valuable clues to encourage cross-religion communications.
}\\[3ex]

Religion is considered as a notable origin of interpersonal relations, as well as an effective and efficient tool to organize a huge number of people towards some challenging targets. At the same time, a believer prefers to make friend with other people of the same faith, and thus people of different faiths tend to form isolated and homogeneous communities \cite{McPherson2001Birds}. Such religious segregation highly influences (usually negatively) culture evolution, economic development, political pattern, and so on \cite{Ruse2006natural,Bokanyi2016Race}. For example, in addition to the prejudice against atheists \cite{Edgell2006Atheists,Gervais2011Do,Gervais2014Everything,Gervais2017Global}, religious segregation results in increasing conflict and prejudice between religions \cite{Atran2012Religious,Adida2010Identifying,Lindley2002Race}.

\begin{figure*}[htp]
  \centering
  \includegraphics[width=0.8\textwidth]{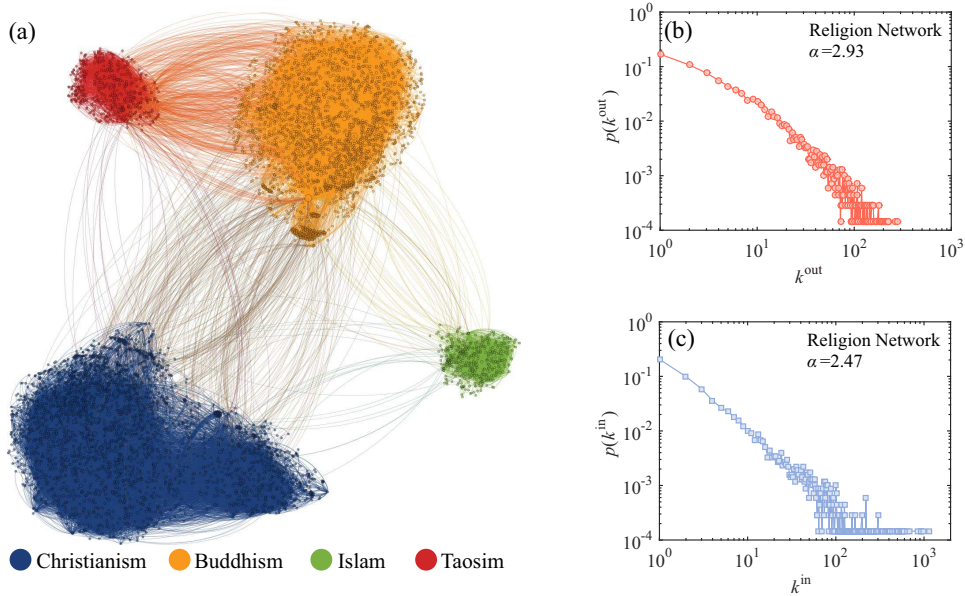}
  \caption{{\bf Structure of the religion network.}  \textbf{(a)}, Structural layout of the network neglecting the directions of links, where blue, orange, green and red nodes denote Christians, Buddhists, Islamists and Taoists, respectively. \textbf{(b)}, The out-degree distribution in a log-log plot, with an estimated power-law exponent $\alpha  \approx 2.93$. \textbf{(c)}, The in-degree distribution in a log-log plot, with an estimated power-law exponent $\alpha  \approx 2.47$.}
    \label{fig1}
\end{figure*}

To quantitatively understand the extent of segregation, we extract a subgraph $G\left( {V,E} \right)$ from \emph{weibo.com}, where $V$ and $E$ denote the sets of nodes and links, respectively. Introduction about \emph{weibo.com} and detailed description of data preparation are shown in Supplementary Note 1. The node set $V$ contains 6875 believers in four major religions in China, including 3153 Christians, 2791 Buddhists, 470 Islamists and 461 Taoists. The link set contains 76678 directed links, and the average degree is 11.15. Figure~\ref{fig1}(a) presents a visual layout of the network, from which one can see clearly that connections inside a religion are dense while connections in-between different religions are much sparser. As shown in Figure~\ref{fig1}(b) and Figure~\ref{fig1}(c), both out-degree and in-degree distributions are approximately power-law, as $p(k) \sim {k^{ - \alpha }}$ where $k$ denotes the degree and $\alpha $ is the power-law exponent. The power-law exponents, estimated by the maximum likelihood method \cite{Clauset2009Power}, are $\alpha  \approx 2.93$ and $\alpha  \approx 2.47$ for out-degree and in-degree distributions, respectively. Induced subgraph for each religion also exhibits scale-free property \cite{Barabasi1999Emergence} (see Supplementary Figure S1), indicating the existence of leaders (with large in-degree) and enthusiasts (with large out-degree).

Neglecting the directions of links, we can obtain an undirected version $G'\left( {V,E'} \right)$ from $G$, where two nodes $i$ and $j$ are connected if either there is a link from $i$ to $j$ or there is a link from $j$ to $i$. In $G'$, there is in total 64712 links. $G'$ displays clustering feature as indicated by its high clustering coefficient \cite{Watts1998Collective} $C = 0.37$, and community structure with a high modularity \cite{Newman2004Finding} $Q = 0.57$ if we directly treat individuals in one religion as one community. However, neither clustering coefficient nor modularity is enough to characterize the aggregation of believers in the same religion or the segregation of believers in different religions, since the former only considers local organization and the latter is very sensitive to the community sizes \cite{Fortunato2007Resolution}. Accordingly, we look into the detailed mixing pattern of the religion network. Denote ${e_{ij}}$ the fraction of links from religion $i$ to religion $j$ ($i,j = 1,2,3,4$), ${a_i} = \sum\limits_j {{e_{ij}}} $ the fraction of links from religion $i$, and ${b_j} = \sum\limits_i {{e_{ij}}} $ the fraction of links pointing to religion $j$, the corresponding mixing matrix is shown in Table 1. Obviously, the religion network is highly assortative, with most links connecting believers in the same religion. In fact, only 1.6\% links are connecting believers of different faiths. We further calculate the assortativity coefficient $r$ \cite{Newman2003Mixing} which lies in [-1,1] with $r=1$ at the perfect assortative mixing (see Methods). The assortativity coefficient of the religion network is surprisingly high, as $r = 0.973$. In comparison, it is even higher than some well-known social networks with remarkable segregation, such as sexual partnerships mixed by races \cite{Catania1992Condom} ($r = 0.621$) and Twitter web of politicians in democratic party and republican party \cite{Conover2011Political} ($r = 0.953$).

\begin{table*}
  \caption{{\bf Mixing matrix of the religion network.} \qquad\qquad\qquad\qquad\qquad\qquad\qquad\qquad\qquad\qquad}
  \centering
  {\begin{tabular*}{14.6cm}{l|cccc|c}
  \hline
  \hline
  \bf Religion & \bf Christianism  & \bf Buddhism  & \bf Islam  & \bf Taoism & \bf ${a_i}$ \\
  \hline
  \bf Christianism & 0.5594303 & 0.0028952 & 0.00010433 & 0.0002739 & 0.56270373 \\
  \bf Buddhism & 0.0017606 & 0.2971778 & 0.000091291 & 0.0048254 & 0.303855091 \\
  \bf Islam & 0.0001956 & 0.0005999 & 0.05676987 & 0.0001304 & 0.05769577 \\
  \bf Taoism & 0.0001695 & 0.0046167 & 0.000013042 & 0.070946 & 0.075745242 \\
  \hline
  \bf ${b_i}$ & 0.561556 & 0.3052896 & 0.056978533 & 0.0761757 &   \\
  \hline
  \hline
  \multicolumn{6}{l}{\multirow{2}{*}{\small The number in $i$th row and $j$th column represent ${e_{ij}}$, the fraction of links from religion $i$ to religion $j$.}}
  \end{tabular*}
  \label{tb1}}
\end{table*}

\begin{table*}
  \caption{{\bf Connecting ratios of the religion network to the null network.}}
  \centering
  {\begin{tabular*}{12cm}{l|cccc}
  \hline
  \hline
  \bf Religion & \bf Christianism  & \bf Buddhism  & \bf Islam  & \bf Taoism  \\
  \hline
  \bf Christianism & 1.7748355 & 0.0169946 & 0.00319361 & 0.0061584 \\
  \bf Buddhism & 0.010329 & 3.1417345 & 0.00538876 & 0.2206321  \\
  \bf Islam & 0.0059429 & 0.0339985 & 18.8441558 & 0.0316456  \\
  \bf Taoism & 0.0039442 & 0.2034483 & 0.00299401 & 12.420091  \\
  \hline
  \hline
  \multicolumn{5}{l}{\multirow{2}{12cm}{\small The number in $i$th row and $j$th column represent the ratio of links from religion $i$ to religion $j$ in the religion network to those in the null network.}}
  \end{tabular*}

  \label{tb2}}
\end{table*}

We further compare the mixing matrix of the religion network $G$ with its randomized counterpart ${G^{\mathrm{null}}}$, which is obtained by the degree-preserved link-rewiring process \cite{Maslov2002Specificity} (see Methods). The mixing matrix of the null network is shown in Supplementary Table S1. We define the connecting ratio from religion $i$ to religion $j$ of $G$ to ${G^{\mathrm{null}}}$ as $\rho _{ij} = e_{ij} / e_{ij}^{\mathrm{null}}$, where $e_{ij}^{\mathrm{null}}$ is the fraction of links from religion $i$ to religion $j$ ($i,j = 1,2,3,4$) in the null network. Table 2 shows such ratios, from which one can observe two remarkable phenomena: (i) Believers statistically tend to connect with others of the same faith as indicated by $\forall i,{\rho _{ii}} > 1$, while Islam and Taoism exhibit the highest level of homophily with ${\rho _{33}} = 18.84$ and ${\rho _{44}} = 12.42$; (ii) The ratios associated with Buddhism, say ${\rho _{2 \bullet }}$ and ${\rho _{ \bullet 2}}$, are all the largest one in corresponding rows and columns excluded the diagonal elements, indicating that Buddhism plays the key role in cross-religion communications in China.

\begin{figure*}[htp]
  \centering
  \includegraphics[width=0.9\textwidth]{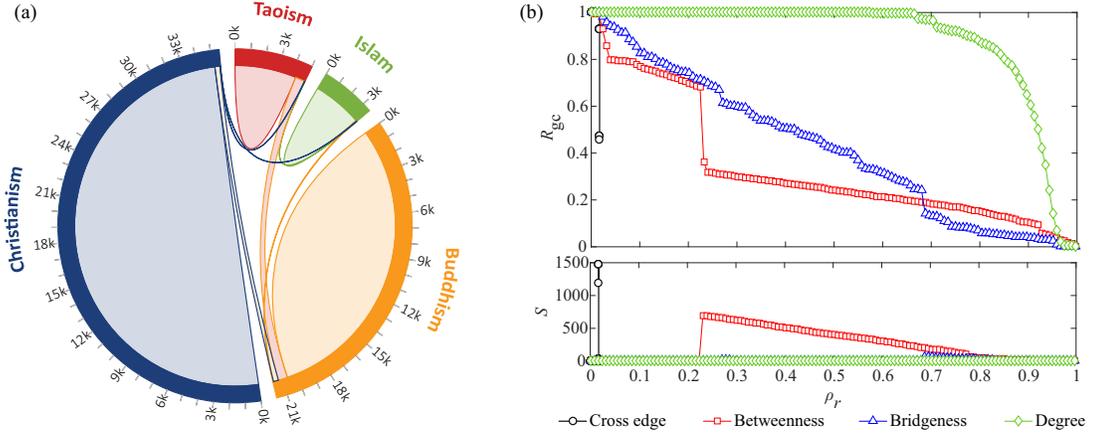}
  \caption{{\bf Quantifying the significance of cross-religion links in maintaining network connectivity. } \textbf{(a)}, Circular plot of links within and between religions, with links' directions being neglected. Inter-religion links are all colored in light colors to emphasize cross-religion links. \textbf{(b)}, Link percolation processes by gradually remove cross-religion links in a random order (black circles), links with largest betweennesses (red squares), links with largest bridgenesses (blue triangles), and links with highest degrees (green diamonds), respectively. The upper and lower plots show the changes of giant component size and normalized susceptibility as the increase of the fraction of removed links.}
    \label{fig2}
\end{figure*}

As indicated by the structural statistics, a tiny number of cross-religion links (i.e., links connecting individuals in different religions, see figure~\ref{fig2}(a) for visualization) play a critical role in maintaining the global connectivity of the religion network. To quantify the significance of cross-religion links, we apply the link percolation dynamics \cite{Onnela2007Structure}, where links are ranked by a certain criterion and then removed one by one in order. For convenience, we consider the undirected version $G'$, wherein there are in total 1124 cross-religion links. The global connectivity is intuitively measured by the ratio of nodes in the giant component (i.e., the largest connected component) to the total number of nodes $N$, denoted by ${R_{GC}}$. Increasing ${\rho _r}$, the fraction of links being removed, the percolation dynamics may come across a phase transition where the network suddenly breaks into many small fragments at the corresponding critical point, accompanied by a sharp drop of ${R_{GC}}$. To precisely locate the critical point $\rho _r^c$, we adopt the normalized susceptibility $\widetilde S{\rm{ = }}\sum\limits_{{\rm{s < }}{{\rm{s}}_{\max }}} {\frac{{{n_s}{s^2}}}{N}} $ \cite{Aharony2003Introduction}, where ${n_s}$ denotes the number of components with size $s$. If there is a percolation phase transition, an obvious peak in the $\widetilde S({\rho _r})$ curve can be observed that corresponds to the critical point $\rho _r^c$, at which the network disintegrates. A set of links whose removal leads to faster decay of ${R_{GC}}$ and smaller value of $\rho _r^c$ is considered to be more significant in maintaining the network connectivity.

We compare the following four methods in identifying significant links for connectivity maintaining: (i) Removing the 1124 cross-religion links in a random order; (ii) Removing links in a descending order of their betweennesses \cite{Girvan2002Community}; (iii) Removing links in a descending order of their bridgenesses \cite{Cheng2010Bridgeness}; (iv) Removing links in a descending order of their degrees \cite{Holme2002Attack}. The explicit definitions of betweenness, bridgeness and degree for an arbitrary link are presented in Methods. As shown in figure~\ref{fig2}(b), as the increasing of ${\rho _r}$, ${R_{GC}}$ decreases much faster when removing cross-religion links first. Remarkable peaks are observed only for the cross-religion links and the largest-betweenness-first method, while the critical point of the former ($\rho _r^c = 1119$) is one order of magnitude smaller than that of the latter ($\rho _r^c = 14956$). In a word, cross-religion links play a remarkably more significant role in maintaining the network connectivity than links with highest betweennesses, bridgenesses or degrees.

\begin{table*}
  \caption{\bf Distribution and average degrees of nodes in different types. \qquad\qquad\qquad\qquad\quad\quad}
  \centering
  \begin{tabular}{p{4.3cm}|p{2.4cm}p{2.4cm}p{2.4cm}p{2.4cm}}
    \hline
    \hline
    & \bf Type1  & \bf Type2  & \bf Type3  & \bf Type4  \\
    & \includegraphics[height=0.45cm]{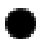} & \includegraphics[height=0.35cm]{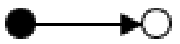} & \includegraphics[height=0.35cm]{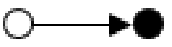} & \includegraphics[height=0.35cm]{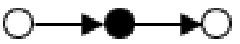} \\
    \hline
    \bf Christianism & 2930 & 153 & 61 (1) & 9 (1) \\
    \bf Buddhism & 2475 & 170 & 78 (24) & 68 (6) \\
    \bf Islam & 417 & 41 & 5 (0) & 7 (0) \\
    \bf Taoism & 271 & 109 & 42 (0) & 39 (1) \\
    \bf Average out-degree & 9.1674 & 25.0085 & 22.7366 & 38.7236 \\
    \bf Average in-degree & 6.7857 & 9.3002 & 134.3978 & 48.2602 \\
    \hline
    \hline
  \end{tabular}
  \\[1.5ex]
  \raggedright
  {\small In each illustration plot, the black node is the ego under consideration and the white node(s) is (are) its neighbor(s). The first four rows show the distribution of nodes of different types in different religions. The numbers in the brackets denote the number of charitable nodes. The average degree of the religion network is 11.15.}
  \label{tb3}
\end{table*}

To uncover the underlying mechanism in the creation of cross-religion links, we classify all nodes into 4 types: (i) nodes not associated with any cross-religion links, (ii) nodes associated with some cross-religion out-links but none of cross-religion in-links, (iii) nodes associated with some cross-religion in-links but none of cross-religion out-links, and (iv) nodes associated with both cross-religion out-links and in-links. Table 3 shows the distribution of nodes of different types in different religions, as well as the average out-degree and in-degree over nodes of different types. Obviously, nodes without any cross-religion links are statistically of smaller degrees than the entire average, while nodes following or being followed by believers of other faiths are generally of higher out-degrees or in-degrees. In particular, the ones being followed by but having not followed believers of other faiths (i.e., Type 3) are usually very popular, with average in-degree more than 10 times larger than the entire average. We further look into the personal descriptions and posted microblogs of nodes of type 3 and type 4. There are in total 309 nodes of type 3 or type 4, each of which has attracted at least one believer from another religion. By content analysis (see Supplementary Note 2), we found 33 charitable nodes who introduce themselves as charity contributors and/or representatives/members of some charity organizations, and have posted a considerable number of charity-related microblogs. To our surprise, such charitable nodes (about 10.7\% of the 309 nodes) have attracted 46.7\% of all cross-religion links, and most charitable nodes (30 of 33) are Buddhists.

In summary, though everybody has observed some evidences about religion segregation in daily life, this paper provides quantitative analysis based on an extracted religion network from \emph{weibo.com.} The extent of networked segregation for different religions, measured by the assortativity coefficient, is even higher than that for different races or different political parties. In fact, to our knowledge, the present religion network exhibits the highest segregation among all previously reported social networks consisted of several classes of people. Among the four religions under consideration, Buddhism plays the most significant role in promoting the cross-religion communications. We still cannot make sure this is a specific phenomenon in China as Buddhism itself is one of a few mainstays of the Chinese culture or a universal phenomenon over the world since the Buddhist doctrines are very inclusive and tolerant. A solid answer to this question asks for more data from \emph{twitter.com} as well as other representative social networks at national level. We have also found that the small-scale religions in China, namely Islam and Taosim, show much higher level of cohesion (see Table 2), which probably reflects a general observation that the subculture group of smaller size usually shows a higher level of homophily \cite{Gelder2007Subcultures}.

A tiny fraction of cross-religion links maintain the global connectivity, whose removal will lead to much faster breakdown of the network in comparison with those links with highest betweennesses or bridgenesses. Therefore we want to understand the underlying reasons of the generation of these cross-religion links. To our surprise, about half links point to charitable nodes. This strong evidence suggests that charity may be a common interest that can stride across the ideological barriers between religions. Accordingly, encouraging and holding charity-related activities, and at the same time inviting participants from different religions, may be an effective method to facilitate cross-religion communications.

In this paper, we demonstrate the effectiveness and validity of the data-driven paradigm in the studies of religious issues, and we believe it will turn to be the mainstream methodology in the near future \cite{Campbell2013,Chen2014}. However, the reported findings just provide a tiny and early step towards the comprehensive landscape of communicating patterns between believers of different faiths. Three open issues are left for further studies. First of all, we would like to test the universality of the present observations based on data from other countries. Secondly, we want to see the evolution of the connecting patterns of religion networks by tracing the temporal data \cite{Holme2012Temporal}. Lastly, it would be interesting to see the role of religious believers in the whole social network, instead of the network containing only believers. This is of particular importance for countries like China where theists are the minority and their social inclusion needs to be promoted.

\section*{Methods}

\noindent \textbf{Assortativity Coefficient.} Assortativity coefficient is used to quantify whether and to which extent links tend to connect nodes in the same type. It is defined as \cite{Newman2003Mixing} $r = \frac{{\sum\nolimits_i {{e_{ii}}}  - \sum\nolimits_i {{a_i}{b_i}} }}{{1 - \sum\nolimits_i {{a_i}{b_i}} }}$ where ${e_{ii}}$, ${a_i}$ and ${b_i}$ are introduced in the main text. In the case of the perfect assortative mixing, all links connecting nodes in the same type, leading to $\sum\nolimits_i {{e_{ii}}}  = 1$ and $r = 1$.

\setlength{\parskip}{0.5\baselineskip}
\noindent \textbf{Degree-preserved link-rewiring process.} This process randomly reshuffles links while keeps the out-degree and in-degree of each node unchanged \cite{Maslov2002Specificity}. At each time step, we randomly select two links $A \to B$ and $C \to D$. If the link $A \to D$ or $C \to B$ exists, we go back to reselect two links, otherwise these two links $A \to B$ and $C \to D$ are replaced by $A \to D$ and $C \to B$. We repeat such operation for sufficiently long time (${10^6}$ steps in this paper) to obtain the randomized counterpart (called null network) of the original network.

\noindent \textbf{Benchmark link centralities.} Betweenness centrality of a link $l$ is the fraction of shortest paths between pairs of nodes passing through $l$ \cite{Girvan2002Community}, say $B{C_l} = \sum\limits_{s,t \in V,s \ne t} {\frac{{\sigma \left( {s,l,t} \right)}}{{\sigma \left( {s,t} \right)}}} $, where $\sigma \left( {s,t} \right)$ is the number of shortest paths between nodes $s$ and $t$, and $\sigma \left( {s,l,t} \right)$ is the number of those paths passing through link $l$. The bridgeness of a link $l$ is defined as ${B_l}{\rm{ = }}{{\sqrt {{S_x}{S_y}} } \mathord{\left/
 {\vphantom {{\sqrt {{S_x}{S_y}} } {{S_l}}}} \right.
 \kern-\nulldelimiterspace} {{S_l}}}$ \cite{Cheng2010Bridgeness}, where $x$ and $y$ are the two endpoints of link $l$. ${S_x}$ and ${S_l}$ are the sizes of the maximum cliques (i.e., complete subgraph) that contain node $x$ and link $l$, respectively. The degree of link $l$ is defined as ${D_l}{\rm{ =  }}{k_x}{k_y}$ \cite{Holme2002Attack}, where ${k_x}$ and ${k_y}$ are the degrees of the two endpoints of $l$.

\end{document}